\documentclass[aip,cha,amsmath,amssymb,reprint,numerical]{revtex4-1}

\usepackage{graphicx}% Include figure files
\usepackage{dcolumn}% Align table columns on decimal point
\usepackage{bm}% bold math
\usepackage{subfigure}
\usepackage{color}
\usepackage{verbatim}

\begin{document}

\title{Edge anisotropy and the geometric perspective on flow networks}

\author{Nora Molkenthin}
\thanks{The first two authors contributed equally to this manuscript.}
\affiliation{Research Domain IV -- Transdisciplinary Concepts and Methods, Potsdam Institute for Climate Impact Research, Telegrafenberg A31, 14473 Potsdam, Germany}
\affiliation{Department of Physics, Humboldt University, Newtonstra{\ss}e 15, 12489 Berlin, Germany}
\affiliation{Network Dynamics Group, Max Planck Institute for Dynamics and Self-Organization, Am Fassberg 17, 37077 G\"ottingen, Germany}
\affiliation{Department of Physics, Technical University of Darmstadt, Hochschulstra{\ss}e 12, 64289 Darmstadt, Germany}

\author{Hannes Kutza}
\affiliation{Research Domain IV -- Transdisciplinary Concepts and Methods, Potsdam Institute for Climate Impact Research, Telegrafenberg A31, 14473 Potsdam, Germany}
\affiliation{Department of Physics, Humboldt University, Newtonstra{\ss}e 15, 12489 Berlin, Germany}

\author{Liubov Tupikina}
\affiliation{Research Domain IV -- Transdisciplinary Concepts and Methods, Potsdam Institute for Climate Impact Research, Telegrafenberg A31, 14473 Potsdam, Germany}
\affiliation{Department of Physics, Humboldt University, Newtonstra{\ss}e 15, 12489 Berlin, Germany}

\author{Norbert Marwan}
\affiliation{Research Domain IV -- Transdisciplinary Concepts and Methods, Potsdam Institute for Climate Impact Research, Telegrafenberg A31, 14473 Potsdam, Germany}

\author{Jonathan F. Donges}
\affiliation{Research Domain I -- Earth System Analysis, Potsdam Institute for Climate Impact Research, Telegrafenberg A31, 14473 Potsdam, Germany}
\affiliation{Stockholm Resilience Centre, Stockholm University, Kr\"aftriket 2B, 11419 Stockholm, Sweden}

\author{Ulrike Feudel}
\affiliation{Institute for Chemistry and Biology of the Marine Environment, Carl von Ossietzky University, Carl-von-Ossietzky-Stra{\ss}e 9, 26129 Oldenburg, Germany}

\author{J\"urgen Kurths}
\affiliation{Research Domain IV -- Transdisciplinary Concepts and Methods, Potsdam Institute for Climate Impact Research, Telegrafenberg A31, 14473 Potsdam, Germany}
\affiliation{Department of Physics, Humboldt University, Newtonstra{\ss}e 15, 12489 Berlin, Germany}
\affiliation{Institute for Complex Systems and Mathematical Biology, University of Aberdeen, Aberdeen, Aberdeen AB243UE, United Kingdom}
\affiliation{Department of Control Theory, Nizhny Novgorod State University, Gagarin Avenue 23, 606950 Nizhny Novgorod, Russian Federation}

\author{Reik V. Donner}
\affiliation{Research Domain IV -- Transdisciplinary Concepts and Methods, Potsdam Institute for Climate Impact Research, Telegrafenberg A31, 14473 Potsdam, Germany}

\date{\today}

\begin{abstract}
Spatial networks have recently attracted great interest in various fields of research. While the traditional network-theoretic viewpoint is commonly restricted to their topological characteristics (often disregarding existing spatial constraints), this work takes a geometric perspective, which considers vertices and edges as objects in a metric space and quantifies the corresponding spatial distribution and alignment. For this purpose, we introduce the concept of edge anisotropy and define a class of measures characterizing the spatial directedness of connections. Specifically, we demonstrate that the local anisotropy of edges incident to a given vertex provides useful information about the local geometry of geophysical flows based on networks constructed from spatio-temporal data, which is complementary to topological characteristics of the same \emph{flow networks}. Taken both structural and geometric viewpoints together can thus assist the identification of underlying flow structures from observations of scalar variables. 
\end{abstract}


\maketitle

\begin{quotation}

Complex networks have recently attracted a rising interest for studying dynamical patterns in geophysical flows like in the atmosphere and ocean. For this purpose, two distinct approaches have been proposed based on either (i) correlations between values of a certain variable measured at different parts of the flow domain (correlation-based flow networks) or (ii) transition probabilities of passively advected tracers between different parts of the fluid domain (Lagrangian flow networks). So far, investigations on both types of flow networks have mostly addressed classical topological network characteristics, disregarding the fact that such networks are naturally embedded in some physical space and, hence, have intrinsic restrictions to their structural organization. In this paper, we introduce a novel concept to obtain a complementary geometric characterization of the local network patterns based on the anisotropy of edge orientations. For two prototypical model systems of different complexity, we demonstrate that the geometric characterization of correlation-based flow networks derived from scalar observables can actually provide additional and useful information contributing to the identification of the underlying flow patterns which are often not directly accessible. In this spirit, the proposed approach provides a prospective diagnostic tool for geophysical as well as technological flows.

\end{quotation}

\section{Introduction} % (fold)
\label{sec:introduction}

During the last years, the application of concepts of complex network analysis has reached a variety of scientific disciplines \cite{Strogatz:2001,Albert:2002,Newman:2003aa,Boccaletti:2006a}. In a growing number of studies, the analyzed networks have been embedded in some physical space \cite{Barthelemy:2011}, which implies that their vertices take well-defined positions and edges describe physical connections (or, more generally, interdependencies) within this space. Examples for such spatial networks can be found in diverse fields such as infrastructures (e.g., road networks, power grids, etc.) \cite{Gastner:2006,Lammer:2006,Chan:2011}, neuronal (brain) networks \cite{Bialonski:2010}, or network representations of the dynamical similarity between climate variations observed at distant points on the globe commonly referred to as (functional) climate networks \cite{Tsonis:2004, Donges:2009a, Donges:2009b, Tsonis:2011, Steinhaeuser:2011,Boers2014,Zhou2015}. 

Due to their embedding in some metric space, spatial networks are not completely described by their \emph{topological} characteristics. By contrast, the \emph{geometric} structure of these systems often needs to be taken into account as well \cite{Heitzig2012,Molkenthin:2014,Wiedermann2015, Tupikina2015}. The latter aspect relates to an entirely different class of spatial network characteristics \cite{Chan:2011}. At the vertex level, the spatial heterogeneity of vertex positions can be quantified by their local density. Regarding the edges, the spatial vertex density and connectivity pattern result in a distinct edge length distribution \cite{Masucci:2009,Chan:2011}. In addition to the latter property, this work proposes characterizing the heterogeneity of the spatial orientations of all edges associated with a given vertex as a complementary aspect. For this purpose, we introduce the concept of \emph{edge anisotropy} as a geometric means to quantitatively describe this feature. In the field of road network analysis, a conceptually related approach, the orientation or trend entropy, has been proposed recently, which also characterizes the heterogeneity of road orientations in physical space \cite{Gudmundsson:2013,Mohajeri:2013,Mohajeri:2014}. Here, we take a more formal approach by describing the anisotropy of edges at both, vertex level and global network scale.

In this work, we explore the potentials of the utilization of edge anisotropy in combination with established topological network characteristics for unveiling the spatial connectivity structure underlying geophysical flow patterns. Specifically, we apply this new concept to characterize the local spatial organization of two-dimensional flow systems using spatially embedded \emph{functional network} representations based on the linear correlations among fields of scalar-valued time series. As illustrative examples, we consider two spatio-temporally discretized flow systems representing (i) advection-diffusion dynamics of temperature in a simple meandering flow and (ii) nutrient concentrations in an advection-reaction-diffusion system of ocean currents in the wake of an island. The thus obtained correlation-based \emph{flow networks} \cite{Molkenthin:2013a} (as well as similar approaches based on Lagrangian dynamics \cite{treml:2008,Rossi:2014}) have numerous potential applications in the field of atmospheric or oceanic flows. Note that the term \emph{flow network} is sometimes also used as a synonym for \emph{transportation networks} in technological applications like power grids. In contrast to that, in this work we exclusively consider the aspect of network representations of flows in physical space as exemplified by geophysical flow patterns or, in a similar way, flows in the phase space of dynamical systems.

The remainder of this paper is organized as follows: Section~\ref{sec:local_anisotropy} introduces new geometric network measures characterizing the anisotropy of edge orientations in physical space. In Sections~\ref{sec:results} and \ref{sec:results2}, we discuss the relationship between the topological and geometric characteristics of the flow networks for the two aforementioned prototypical problems constructed in different ways. We compare the spatial patterns of several classical (topological) vertex-based network characteristics and local edge anisotropy in order to demonstrate that the latter adds a new aspect to the network characterization. Our main results are summarized and further discussed in Section~\ref{sec:conclusion_outlook}.

\section{Anisotropy in spatial networks}
\label{sec:local_anisotropy}

\subsection{Preliminaries}

Consider a network with a vertex set $V$ and an edge set $E\subseteq V\times V$, where $N=|V|$ denotes the overall number of vertices. The connectivity pattern of this network is described by its (unweighted) adjacency matrix $\mathbf{A}=(a_{mn})$ with elements
\begin{equation}
	a_{mn} = \left\{
	\begin{array}{ll}
		1, & \mbox{ iff}\ \{m,n\} \in E \\
		0, & \mbox{ else,}
	\end{array}		\right.
\end{equation}
\noindent
where $\{m,n\}$ denotes an edge from vertex $m$ to vertex $n$. The degree $k_m$ of vertex $m$, i.e., the number of edges adjacent to this vertex, is then given by summing up all entries in the $m$-th row of $\mathbf{A}$, $k_m=\sum_{n=1}^N a_{mn}$.

Let us now suppose that the vertices are associated with well-defined Cartesian coordinates $\vec{x}_m=(x_m^{(1)},\dots,x_m^{(d)})$ in a $d$-dimensional Euclidean space. Here, the coordinates are provided with respect to an arbitrarily chosen origin. For a given vertex at $\vec{x}_m$, each adjacent edge $\{m,n\}$ or $\{n,m\}$ connecting $m$ (in outward or inward direction, respectively) with another vertex $n$ located at $\vec{x}_n$ can be fully characterized by its length $l_{mn}=\|\vec{x}_n-\vec{x}_m\|$ and its spatial orientation described by the unit vector $\vec{e}_{mn}=l_{mn}^{-1}(\vec{x}_n-\vec{x}_m)$.

\subsection{Local anisotropy}

The \emph{local anisotropy} of edge orientations (or, for short, local (edge) anisotropy) $R_m$ is a geometric network measure which characterizes the heterogeneity of orientations of all edges adjacent to a given vertex and, hence, the spatial directedness of this vertex' connectivity. We emphasize that this aspect is potentially relevant for transport and distribution networks, where it is commonly advantageous to locate the sources of material flows (logistic hubs) in the center of the area to be served rather than at its periphery to minimize transportation costs~\cite{Kreher:2012}. This idea calls for a generally high degree of isotropy of the transportation routes at the distributor node, which should be reflected by a low value of our anisotropy measure.

For defining the local anisotropy of a given vertex $m$ in an unweighted and undirected ($a_{mn}=a_{nm}\; \forall\, m,n=1,\dots,N$) two-dimensional spatial network, we consider the Rayleigh measure computed from the orientations (measured in terms of the unit vectors $\vec{e}_{mn}$) of all edges adjacent to $m$,
\begin{equation}
R_m%=\frac{1}{k_m} \left\|\sum_{n=1}^N a_{mn} \frac{\vec{x}_{n}-\vec{x}_m}{\|\vec{x}_{n}-\vec{x}_m\|}\right\|
=\frac{1}{k_m} \left\|\sum_{n=1}^N a_{mn} \vec{e}_{mn} \right\| \in [0,1].
\label{eq:rayleigh}
\end{equation}
\noindent
Practically, $R_m$ projects the vectors $\vec{e}_{mn}$ describing the spatial orientations of all existing edges adjacent to vertex $m$ onto the $d$-dimensional unit sphere, thereby taking their \emph{spatial orientation} (but \emph{not} their length) into account, and calculates the modulus of the vector sum of all corresponding unit vectors. The normalization by degree $k_m$ ensures $R_m \in [0,1]$ for an angular distribution between maximally unfocused (isotropic, $R_m=0$) and maximally focused (anisotropic, $R_m=1$) edge directions. 

\begin{figure}
\centering
	\includegraphics[width = 0.75\columnwidth]{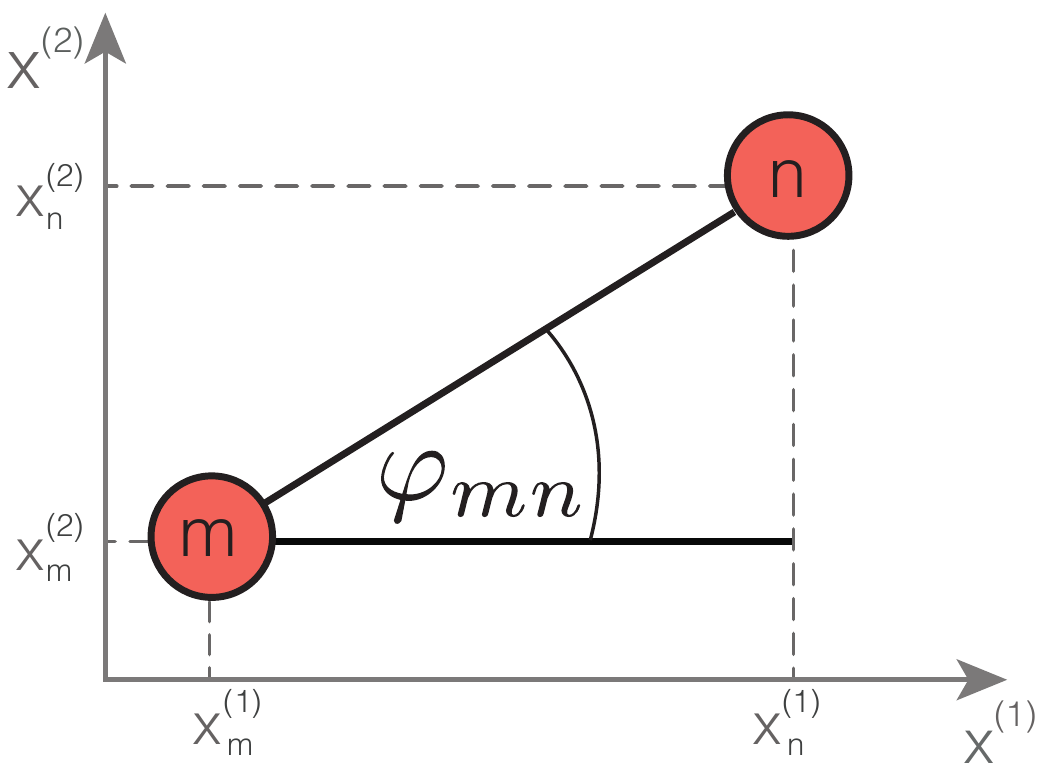}
	\caption{Schematic definition of the (planar) edge angle $\varphi_{mn}$ between two vertices $m$ and $m$ with given two-dimensional Cartesian coordinates $x_{m}^{(1,2)}$, $x_{n}^{(1,2)}$, which are connected by an undirected edge (as common for the correlation-based flow networks considered later in this work). Without loss of generality, $x^{(1)}$ is used here as the axis of reference for defining $\varphi_{mn}$ for all edges $\{m,n\}$.}
	\label{illustration_anisotropy_angle}
\end{figure}

In order to illustrate this concept, let us consider the special case of $d=2$,
providing a simplified model of geographical space by neglecting curvature due to the approximately spherical shape of the Earth's surface. In this setting, we can assign a Euclidean angle $\varphi_{mn}$ with respect to an arbitrary reference axis (here, we use the $x^{(1)}$ axis without loss of generality) by setting (cf.~Fig.~\ref{illustration_anisotropy_angle})\footnote{Practically, $\varphi_{mn}$ is computed numerically using the \texttt{atan2} function in most common statistical software packages yielding \texttt{atan2}$(x_n^{(1)}-x_m^{(1)},x_n^{(2)}-x_m^{(2)}) \in [-\pi,+\pi]$.}
\begin{equation}
\varphi_{mn} = \arctan\left(\frac{x^{(2)}_n - x^{(2)}_m}{x^{(1)}_n - x^{(1)}_m}\right).
\end{equation}
\noindent
This corresponds to replacing the former Cartesian coordinates by local polar coordinates centered at $\vec{x}_m$, yielding 
\begin{equation}
\vec{x}_{n}-\vec{x}_m = l_{mn}\,\left(\begin{array}{c}\cos\varphi_{mn}\\ \sin\varphi_{mn}\end{array}\right),
\end{equation} 
\noindent 
In this case, we can rewrite Eq.~(\ref{eq:rayleigh}) as 
\begin{equation}
R_m = \frac{1}{k_m} \left|\sum_{n=1}^{N} a_{mn} e^{i \varphi_{mn}}\right|.
\label{eq:mrl}
\end{equation}

A generalization to weighted networks is easily obtained by replacing the binary adjacency matrix $\mathbf{A}$ by the edge weight matrix $\mathbf{W} = (w_{mn})$ ($w_{mn} \in \mathbb{R}^+\; \forall\, m,n$). In this case, a proper normalization factor is given by the associated vertex strength $s_m = \sum_{n=1}^N w_{mn}$ replacing $k_m$, so that
\begin{equation}
	R_m = \frac{1}{s_m} \left\|\sum_{n=1}^{N} w_{mn} \vec{e}_{mn}\right\| \in[0,1].
\end{equation}
\noindent

A corresponding definition for directed yet unweighted networks is based on replacing $k_m$ by the in- and out-degrees $k_m^{in}$, $k_m^{out}$ and taking the sum either over incoming or outgoing edges, yielding the in- and out-anisotropies, respectively, 
\begin{eqnarray}
	R_m^{in} &=& \frac{1}{k_m^{in}} \left\|\sum_{n=1}^{N} a_{nm} \vec{e}_{mn}\right\| \; \mbox{with} \ k_m^{in} = \sum_n a_{nm},\\
	R_m^{out} &=& \frac{1}{k_m^{out}} \left\|\sum_{n=1}^{N} a_{mn} \vec{e}_{mn}\right\| \; \mbox{with} \ k_m^{out} = \sum_n a_{mn}.
\end{eqnarray}
\noindent
As for the undirected case, the generalization to weighted networks is obtained by replacing $a_{mn}$ by $w_{mn}$ and the in- and out-degrees by the in- and out-strengths $s_m^{in}=\sum_n w_{nm}$ and $s_m^{out}=\sum_n w_{mn}$, respectively.

\subsection{Related measures}

The proposed concept of anisotropy of edge orientations has several possible extensions as well as linkages to related characteristics. Although the examples discussed in this paper will focus exclusively on the local anisotropy in networks characterizing spatial flow patterns, we are confident that some of these extensions are of potential interest for studying general spatial networks. In the following, we will briefly discuss some of these aspects.

At the local (vertex) scale, anisotropy can be characterized not just by $R_m$, but also a variety of other measures. For $d=2$, one immediately recognizes the formal analogy between Eq.~(\ref{eq:mrl}) and the mean resultant length commonly used for quantifying the degree of phase synchronization between two mutually coupled oscillators in terms of differences between their respective phase dynamics on the unit circle \cite{Pikovsky2001}. Accordingly, corresponding alternative characteristics serving the same purpose include the circular standard deviation and Shannon entropy of $\varphi_{mn}$ for a given $m$. In comparison with the Rayleigh measure, both have certain disadvantages. On the one hand, the standard deviation is commonly not normalized. On the other hand, the Shannon entropy can be normalized, but its estimation relies on some binning of the interval $[-\pi,\pi]$ and can therefore only be properly performed in case of sufficiently high vertex degrees.

Going to the global network scale, one common approach is defining scalar network characteristics by taking some mean value over an associated vertex measure. A prominent example for this strategy is the global clustering coefficient defined as the arithmetic mean of the local clustering coefficients of all vertices in a network \cite{Watts:1998}. In a similar way, one possibility to characterize the edge anisotropy of the network at a global scale is taking the \emph{mean local anisotropy}
\begin{equation}
\overline{R}=\frac{1}{N}\sum_{m=1}^N R_m.
\end{equation}
\noindent
We note that this definition gives different effective weight to edges associated with vertices of different degrees. To see this, recall that each $R_m$ is defined as a sum over terms related to different edges adjacent to $m$, which contribute to $R_m$ equally with the same weight $1/k_m$ (in the case of unweighted networks, otherwise $1/s_m$). Thus, if an edge connects two vertices with low degree, it contributes much stronger to $\overline{R}$ than edges connecting vertices with high degrees. In order to correct for this effect, we propose studying the \emph{global anisotropy}
\begin{equation}
R=\frac{1}{\sum_{m,n=1}^N a_{mn}}\left\|\sum_{m,n=1}^N a_{mn}\vec{e}_{mn}\right\|
\label{eq:rglob}
\end{equation}
\noindent
instead, where each edge has the same weight\footnote{In Eq.~(\ref{eq:rglob}), every edge is counted twice in the vector sum, which is corrected for by the considered normalization factor.}. Drawing upon the analogy to the network's clustering properties, $\overline{R}$ and $R$ take the roles of the Watts-Strogatz and Barrat-Weight definitions of the clustering coefficient, respectively \cite{Watts:1998,Barrat:2000}, the latter of which is also referred to as network transitivity in the literature \cite{Boccaletti:2006a}.

As for the local measures, it is also possible to replace the mean resultant length by the circular standard deviation or Shannon entropy of edge angles $\varphi_{mn}$ in the definition of global anisotropy properties. While this replacement still suffers from the same conceptual problems as the corresponding vertex characteristics when considering the mean local properties, a corresponding modification of the global anisotropy concept relieves the previous problem of too small sample sizes (arising especially in the case of sparse networks). In case of the Shannon entropy, a corresponding measure (referred to as \emph{trend entropy}~\cite{Gudmundsson:2013,Mohajeri:2013,Mohajeri:2014}) has been recently used for the analysis of street network patterns.

Under general conditions, we emphasize that it can also be interesting to consider the \emph{direction} of the vector sum of all unit vectors characterizing the edges adjacent to a given vertex,
\begin{equation}
\vec{e}_m=\frac{\sum_{n=1}^N a_{mn} \vec{e}_{mn}}{\left\|\sum_{n=1}^N a_{mn} \vec{e}_{mn}\right\|}
\end{equation} 
\noindent
(which straightforward generalizations for weighted and/or directed spatial networks) instead of just the modulus $R_m$, especially for the purpose of visualization of flows on a given spatial network. In a similar spirit, in certain applications the length and orientation of the resultant vector $\vec{r}_m=\sum_{n=1}^N a_{mn}\, \|\vec{x}_n-\vec{x}_m\|$ might be of interest as well. However, this directional aspect (which has been recently studied using a conceptually related measure in the context of regional climate network presentations to unveil the spatial structures of heavy precipitation events \cite{Rheinwalt2015}) is beyond the scope of this work focusing on quantitative rather than qualitative network characterization.

\section{Example 1: Advection-diffusion dynamics of temperature in a stationary flow} \label{sec:results} %"meandering" sounds less general

In the following, we will illustrate how a combination between classical (topological) network characteristics and the new concept of local anisotropy can help gaining additional understanding about the structural organization of spatially embedded systems. For this purpose, we study flow networks constructed from correlations among fields of scalar observables for two prototypical flow systems with different levels of structural complexity. Note that these networks do not characterize the mean state of the flow under study, but the spatial interdependence between fluctuations superimposed to this baseline flow.

\subsection{Discretized advection-diffusion systems}

In order to describe the temperature dynamics in a fluid moving with a given two-dimensional velocity field $\vec{v}(\vec{x})$ (an extension to three-dimensional flows is possible, but shall not be further studied hereafter), we consider the classical advection-diffusion equation (ADE) \cite{Lebon2008}
\begin{equation}
\frac{\partial T}{\partial t} = \kappa \Delta T - \vec{v}(\vec{x})\nabla T +   D\xi (\vec{x},t)
\label{ade}
\end{equation}
\noindent
for an incompressible fluid, where $\kappa$ is the diffusion coefficient and $D$ the intensity of noise superimposed to the deterministic flow equations. The temperature outside the considered flow domain is fixed at zero, implying that the thermal conditions of the medium outside this domain do not affect those inside. For simplicity, we consider here dimensionless variables, a time-independent velocity field and uncorrelated Gaussian white noise with zero mean, unit variance and no spatial or temporal correlations (i.e., $\langle \xi (\vec{x},t)\xi (\vec{y},t')\rangle=\delta(\vec{x}-\vec{y})\delta(t-t')$). Furthermore, we do not take possible effects of temperature variations on the velocity field $\vec{v}(\vec{x})$ into account, but leave the latter independent of temperature.

For constructing a flow network based on the scalar temperature field $T(\vec{x},t)$, we first consider a spatio-temporal discretization of Eq.~(\ref{ade}) 
without the stochastic diffusion term, using an Euler scheme and a regular square lattice with spatial resolution $\Delta x$~\cite{toral2014}. 
The corresponding discretization parameters $\Delta x$ and $\Delta t$ are chosen such to fulfill the Courant-Friedrichs-Lewy conditions~\cite{press1988} 
to ensure the stability of the discretization scheme. Letting $i$ and $j$ denote the grid indices in $x$- and $y$-direction, respectively, and $\vec{v}_{i,j}=(v^x_{i,j}, v^y_{i,j})$ 
being the fluid velocity at a given grid point, the discretized ADE at time step $t'$ takes the form
\begin{equation}
\begin{split}
&T_{i,j}(t'+1)= (1-4\kappa) T_{i,j}(t') +\\
& \quad +(\kappa-v^x_{i,j}/2) T_{i+1,j}(t') + (\kappa-v^y_{i,j}/2) T_{i,j+1}(t') +\\
& \quad + (\kappa+v^x_{i,j}/2) T_{i-1,j}(t') + (\kappa+v^y_{i,j}/2) T_{i,j-1}(t').
\label{ade:discrete}
\end{split}
\end{equation}
\noindent
Due to the considered boundary conditions as described above, some of the coefficients in Eq.~(\ref{ade:discrete}) corresponding to vertices at the boundary of the considered spatial domain take zero values. We emphasize that the employed discretization scheme is intentionally kept very simple and does not consider the more complex structure of schemes commonly used in geophysical or technological flow dynamics. Specifically, in the present example, both scalar field (temperature) and vector field (flow velocity) are calculated at the same grid.

Taken together, the dynamics of the discretized scalar temperature field $\vec{T}(t')=(T_1(t'),\dots,T_N(t'))$ ($t'=0,\dots,t-1$) under the action of the underlying flow can be approximated by a linear recursive equation with additive noise,
\begin{equation}
\vec{T}(t'+1)= \mathbf{P} \vec{T}(t') + s \vec{\epsilon}(t'),
\label{recur}
\end{equation}
where $\mathbf{P}$ is a matrix approximating the time-evolution operator of the advection-diffusion process, and $\vec{\epsilon}(t')$ is a vector of independent Gaussian random variables of zero mean and unit variance, uncorrelated at different time steps. $s=\sqrt{D\Delta t/\Delta x^2}$ denotes the intensity of the discretized noise \cite{toral2014}, which we set to $1$ in the following. Neglecting the stochastic term in this vector autoregressive model equation, we can easily obtain the matrix elements of $\mathbf{P}$ from the coefficients of the discretized ADE in Eq.~(\ref{ade:discrete}). In particular, the time-evolution matrix $\mathbf{P}$ can be constructed for each given stationary velocity field. In this case, $\mathbf{P}$ is a time-independent matrix itself. In addition, Eq.~(\ref{recur}) can be extended to incorporate secondary effects such as an external perturbation in some part of the domain of interest (see below).

Following the previous considerations, Eq.~(\ref{recur}) allows us to generate a field of time series from an initial temperature field $\vec{T}(0)$, which we assume here to be zero everywhere without loss of generality since the resulting asymptotic dynamics is independent of the initial conditions \cite{Tupikina2015}. Specifically, solving Eq.~(\ref{recur}) for $t$ time steps yields a vector moving-average process of order $t$ describing the temperature dynamics at each grid point depending on the imposed noise,
\begin{equation}
\vec{T}(t) = \sum_{k=0}^{t-1}\mathbf{P}^{t-1-k} \vec{\epsilon} (k),
\end{equation}
\noindent
where the advective dynamics is fully encoded in the matrix elements of $\mathbf{P}$.

The latter observation is useful for obtaining an analytical representation of the covariance matrix $\mathbf{\Gamma}$ between the temperature evolution at all grid points from the discretized time-evolution operator. Note that the covariance between two time series with zero mean is generally defined as the sum over the products of all simultaneously observed values of the two time series, i.e., the scalar product of the two vectors. In a similar spirit, the covariance matrix taking all grid points into account is defined as the sum over the respective tensor products of concurrent temperature values. Due to the imposed noise, we take the expectation value yielding
\begin{equation}
\begin{split}
\mathbf{\Gamma}&= \left< \sum_{t'=0}^t \vec{T}(t') \otimes \vec{T}(t') \right> \\
&= \sum_{t'=0}^t \sum_{k=0}^{t'-1} \sum_{k'=0}^{t'-1} \left< \mathbf{P}^{t'-1-k}\vec{\epsilon}(k) \otimes \mathbf{P}^{t'-1-k'}\vec{\epsilon}(k') \right>.
\end{split}
\end{equation}
\noindent
By evaluating the expectation values in the inner sums, the latter equation can be conveniently reformulated as
\begin{equation}
 \mathbf{\Gamma} =\sum_{t'=0}^{t}\sum_{k=0}^{t'-1}(\mathbf{P}{\mathbf{P}}^T)^{t'-1-k},
 \label{noise3}
\end{equation}
\noindent
which converges if $|\lambda_{max}|\leq 1$ with $\lambda_{max}$ being the eigenvalue of $\mathbf{P}$ with the largest modulus. By normalization with respect to the diagonal elements of $\mathbf{\Gamma}$, one easily obtains the associated correlation matrix $\mathbf{C}$. In the latter, the entry $C_{mn}$ denotes the lag-zero correlation coefficient between the time series $T_m(t)$ and $T_n(t)$.

\subsection{Meandering flow model}

\begin{figure*}
	\includegraphics[width = 0.7\textwidth]{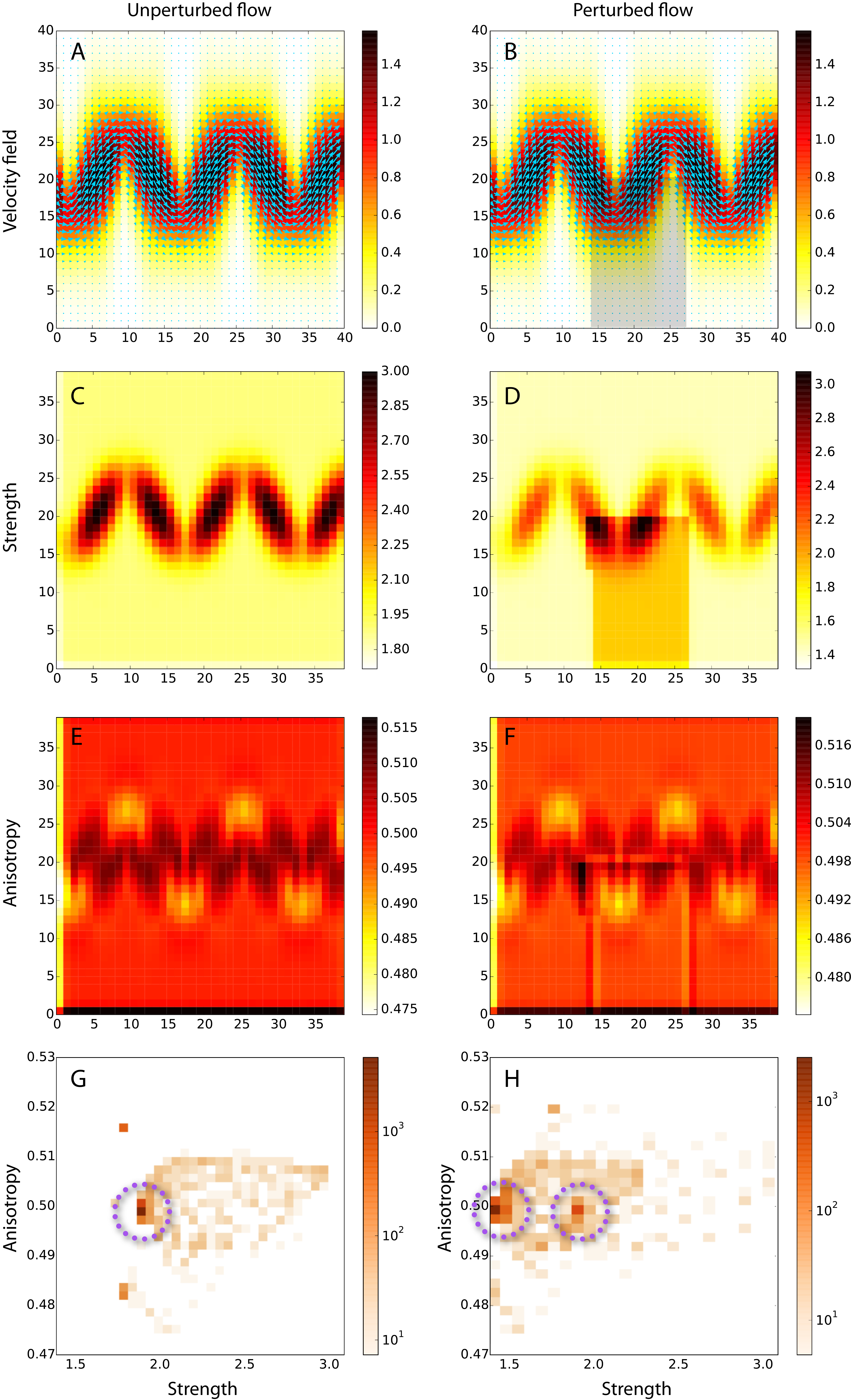}
	\caption{(Color online) (A,B) Velocity field of the meandering flow example (colors indicating the velocity modulus prior to rescaling by the factor $v_0$, arrows the flow direction) without (A) and with (B) a perturbed region (shaded area, see the main text for further details). In addition, the vertex strengths (C,D) and local anisotropies (E,F) of the associated flow networks are shown for the unperturbed (C,E) and perturbed case (D,F). Panels (G,H) display the two-dimensional histograms of local anisotropy and vertex strength (colors indicating the frequency of all combinations between the respective values). Note that the corresponding axes show the values of both network measures, which is different from panels (A)-(F) where the spatial positions of vertices are indicated. Notable groups of many vertices with approximately the same strength and anisotropy (see discussion in the main text) are additionally highlighted by dashed circles.}
	\label{fig:mf}
\end{figure*}

As a specific example, we consider a stationary velocity field representing a variant of 
the classical two-dimensional meandering flow model~\cite{cencini1999mixing, bower1991simple, Samelson1992, Raynal2006, lopez2001} (Fig.~\ref{fig:mf}A)
\begin{equation}
\vec{v}(x,y)=v_0 \left(-\frac{\partial\psi}{\partial y},\frac{\partial\psi}{\partial x}\right)
\end{equation}
\noindent
with the velocity magnitude $v_0$ and the stream function
\begin{equation}
\Psi(x,y) = 1- \tanh \left[\frac{y- \sin 2x}{\cos(\arctan(\cos 2x ))}\right],
\label{psi}
\end{equation}
where $v_0=0.2$ has been chosen such that the maximum velocity at all grid points does not exceed $0.5$. 

For the discretization scheme, we use $\Delta t=\Delta x=1$ and a square lattice of $40\times 40$ grid points, yielding $N=1,600$ vertices of the flow network. In order to comply with the Courant-Friedrichs-Lewy criteria $\kappa\Delta t/(\Delta x)^2\leq 1$ and $\Delta t \cdot \max_{\vec{x}} |\vec{v}(\vec{x})|\leq \Delta x$, we consider a dimensionless thermal diffusion coefficient of $\kappa=0.01$. The integration time of the system was taken as $t=20$. 

In addition to the unperturbed flow, we consider the case where some part of the flow domain is affected by an external perturbation. For example, consider a region where the fluid has a larger heat capacity or is "externally heated" (see Fig.~\ref{fig:mf}B). 
This perturbation is realized by adding an additional term to the diagonal elements of $\mathbf{P}$, yielding
\begin{equation}
\mathbf{P}'=\mathbf{P}+\mathbf{H}.
\end{equation}
\noindent
Here, $\mathbf{H}$ is a diagonal matrix with entries $h_{mm}>0$ in externally heated regions and $h_{mm}=0$ elsewhere. For the sake of simplicity, we just take uniform $h_{mm}=0.03$ for all grid points $m$ inside the perturbed region highlighted in Fig.~\ref{fig:mf}B, leaving the detailed exploration of other settings a subject for future work. After a possibly necessary renormalization to ensure convergence, $\mathbf{P}'$ again serves as a discretized time-evolution operator of the ADE process under the considered stationary flow.

\subsection{Flow network construction}

The correlation matrix of the discretized ADE system as derived above can be directly re-interpreted as the weight matrix of an undirected weighted flow network by setting
\begin{equation}
w_{mn}=C_{mn}-\delta_{mn},
\label{def:network_weighted}
\end{equation}
\noindent
where $\delta_{mn}$ is Kronecker's delta used for removing the trivial correlation of each grid point with itself. Here, the considered grid points represent the vertices of the constructed network. Given the dynamics of $\vec{T}$ and the associated estimate of $\mathbf{C}$ based on $\mathbf{P}$, this network directly represents the linear correlations among the temperature variations observed at all grid points $m=1,\dots,N$.

Before further analysis, it is important to highlight the practical interpretation of the thus constructed flow network. For the considered homogeneous initial conditions $T_m(0)=0$ $\forall\, m\in V$ in the absence of additional gradual perturbation $\mathbf{H}$, the mean temperature field will be described by a homogeneous equilibrium. However, the correlation-based flow networks do not characterize the corresponding mean state, but the spatial interdependencies between fluctuations superimposed to this equilibrium. Both diffusive and advective terms transport the noisy input signal, implying that the fluctuations at nearby spatial locations are correlated even in the case of independent noise. However, the corresponding correlations can be expected to be no simple functions of the spatial distance between two vertices, but also reflect the structure of the underlying velocity field. In this spirit, the correlation-based flow network should contain information on both diffusive and advective structures (and, hence, the considered flow pattern). In the following, we will examine qualitatively how different flow network characteristics can be used for inferring the corresponding information.

\subsection{Flow network characteristics}

In order to study the spatial connectivity patterns of the correlation-based flow network, we first consider the individual vertex strengths $s_m$. For purely ad\-vec\-tion-diffusion 
based systems, it was shown previously \cite{Molkenthin:2013a} that a high degree commonly indicates high velocity. This observation is further supported by our example 
(to see this, compare the high-velocity regions in Fig.~\ref{fig:mf}A,B with the locations of vertices with high strength in Fig.~\ref{fig:mf}C,D). However, Fig.~\ref{fig:mf}D additionally 
reveals that the vertex strength can also be elevated due to a common trend at all vertices in a certain area within the discretized scalar field $\vec{T}$ increasing the correlation coefficients among these vertices. Similar observations have been made recently for the degree fields of climate networks constructed from surface air temperature anomalies in the presence of volcanic eruptions or strong El Ni\~no episodes \cite{Radebach:2013}.

According to the latter result, it is evident that another network measure is necessary to distinguish between the effects of external heating and high velocity. 
As a potential candidate, the local anisotropy of the system is shown in Fig.~\ref{fig:mf}E,F. As a prominent feature of the flow network geometry, 
we observe that regions with high fluid velocity are commonly surrounded by areas with elevated anisotropy values whenever the local flow pattern is straight. 
This feature can be explained by strongly correlated grid points being aligned with the flow, i.e., spatial positions within the flow domain which experience temporal variations of the local temperature at the same time due to the advection. 
As an exception, in the areas with the highest stationary velocity (and highest vertex strength), the local anisotropy is reduced in comparison with regions slightly apart from the center of the flow. We relate this to the fact that a high vertex strength $s_m$ implies a large number of vertices being strongly connected, which cannot all be aligned linearly.

In contrast to the areas with high fluid velocity and straight flow geometry, if the (stationary) flow is fast but curved (especially at the turning points of the meanders), 
the locally linear geometric alignment of the most strongly correlated vertices is relieved, resulting in lower anisotropy values. In fact, we observe the lowest anisotropy values 
among all vertices in those regions where the flow takes a sharp turn. Moreover, the local anisotropy values are also slightly reduced in regions of slow flow, 
where diffusive heat transport has a larger relative importance in comparison with advective one than under fast flow conditions. 
Hence, correlations among temperature fluctuations can also be relevant in directions perpendicular to the flow.

Another notable finding is the absence of marked differences between the cases without and with external heating in the area where the perturbation is applied -- 
with the exception of the boundaries of the perturbed domain that still deviate from the almost constant background anisotropy values. 
Hence, in combination with the vertex strength, 
local anisotropy can be used to distinguish high velocity areas from regions with common temperature trends due to external forcing. 
This result suggests that anisotropy and related geometric network properties are potentially useful tools for the analysis of correlation-based flow networks, 
but also other spatial networks like climate networks.

In order to gain further understanding of the complex interplay between vertex strength and anisotropy, Fig.~\ref{fig:mf}G,H shows scatter plots between both characteristics for the cases without and with external perturbation. In the unperturbed case (Fig.~\ref{fig:mf}G), we find a large group of vertices corresponding 
to the large area of slow stationary flow, which are characterized by relatively low strength and medium anisotropy values (dashed circle). Vertices from the faster flow area (i.e., with larger strength) show a relatively broad range of anisotropy values, depending on whether the respective vertices are located at a turn (low $R_m$) or a straight segment of the flow pattern (high $R_m$). In the case of the perturbed flow (Fig.~\ref{fig:mf}H), the network maintains these basic features. However, much of the empirical distribution of vertex strengths is shifted towards lower values, reflecting the fact that the correlations between vertices in the perturbed region are elevated due to a common trend, while those of the remaining vertices compensate for this effect by showing generally lower strengths than in the unperturbed case (as also shown in Fig.~\ref{fig:mf}C,D). Specifically, the existence of a homogeneously perturbed region results in a second group of many vertices with approximately the same strength and anisotropy (right dashed circle in Fig.~\ref{fig:mf}H), which corresponds to vertices at grid points with low velocity modulus but external heating. In turn, the original group of vertices with approximately the same network characteristics (dashed circle in Fig.~\ref{fig:mf}G) is shifted towards lower vertex strengths while maintaining their anisotropy values (left dashed circle in Fig.~\ref{fig:mf}H). Thus, the considered perturbation prominently influences the distribution of vertex strengths (i.e., a topological network measure) while retaining most of the anisotropy values (i.e., a geometric characteristic). Future studies should address the question whether similar observations also apply to other flow network characteristics.

\section{Example 2: Nutrient dynamics in a complex advection-reaction-diffusion system} \label{sec:results2}

The previously discussed example has been characterized by a relatively basic flow pattern. In order to demonstrate that the concept of edge anisotropy is also useful for the characterization of more complex flows, in the following we consider a paradigmatic model of a marine current system originally motivated by particle transport in the wake of the Canary Islands. Here, the dynamics of interest is described by an advection-reaction-diffusion (ARD) system, which was already investigated in depth in previous studies~\cite{Sandulescu:2006a,Bastine:2010a}. In the considered setting, a biological model of a marine food web containing available nutrients and different plankton populations is driven by the mesoscale hydrodynamical flow structures in the region. The velocity field is prescribed by a time-periodic stream function and consists of a main background flow, an upwelling region with an Ekman drift and a von K\'{a}rm\'{a}n vortex street. The vortices emerge in the wake of the island due to its role as a major obstacle to the flow and display a mutual phase difference of half the period of the stream function. Unlike in the simple example in Section~\ref{sec:results}, the nutrient concentrations are modeled on a regular Eulerian grid, while the flow dynamics is obtained by means of a Lagrangian approach.

\begin{figure}
	\includegraphics[width = 1.0\columnwidth]{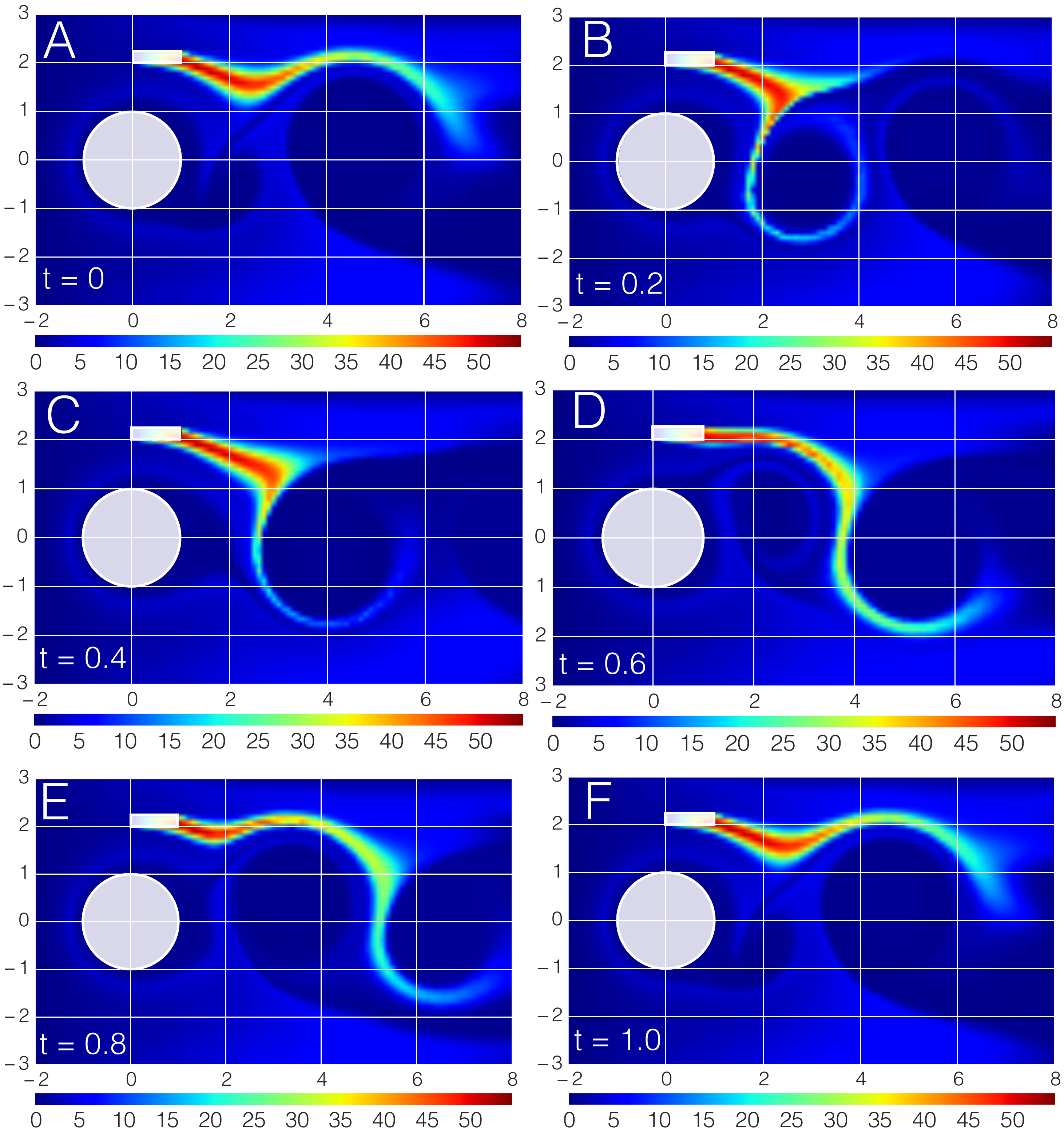}
	\caption{(Color online) Exemplary snapshots of nutrient transport dynamics. The six panels (A-F) depict successive states of nutrient concentrations (colors, arbitrary units) during one period ($T_C=10$ time steps) of the periodic time evolution of the flow. Time is measured in units of $T_C$. The island and the upwelling region are represented by the circle and rectangle, respectively.}
	\label{fig:plankton_sample}
\end{figure}

\begin{figure}
	\includegraphics[width = 1.0\columnwidth]{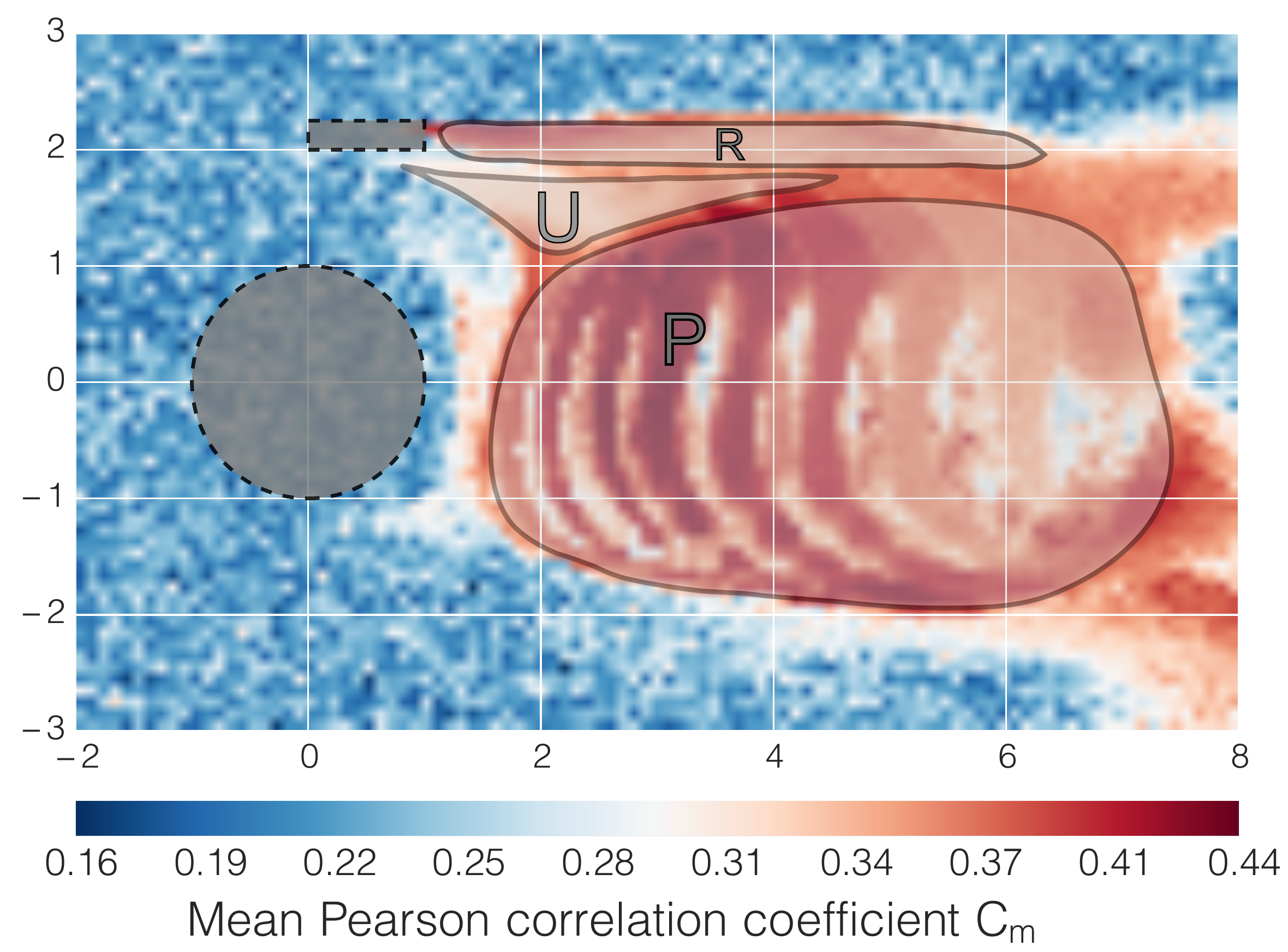}
	\caption{(Color online) Spatial patterns of the mean Pearson correlation (Eq.~\ref{eq:ccflag}) of each grid point in the complex flow pattern with all other grid points, $C_m = N^{-1} \sum_{n=1}^N C_{mn}$. The three marked regions correspond to the surrounding of the upwelling region (U), the ribbon (R) and the plumes region (P). The multiplicity of apparently discrete ribs with high mean Pearson correlation in the plume region arises due to the discrete sampling of the considered nutrient concentrations in time in combination with the imposed periodicity of the flow.}
	\label{fig:pearson_spatial}
\end{figure}

Figure~\ref{fig:plankton_sample} displays the nutrient concentrations (replacing the temperature field of the previous example as a scalar observable) for different phases during one full period of the flow. The vortices in the wake of the island periodically detach with alternating signs of rotation and then travel along the main flow. It was shown \cite{Bastine:2010a} that, depending on the vortex strength, the vortex patterns can either prohibit or permit transport of nutrients and plankton across the wake. In this study, we restrict our attention to the simulated nutrient time series $\hat x_m(t) (t=1,..,T; m=1,..,N)$ as tracers that could be used for the reconstruction of the (possibly unknown) physical flow in the complex hydrodynamical system. Specifically, there are several distinct spatial domains associated with different transport regimes, primarily including the area of and around the upwelling region (U) in the upper left part of the study area, the central plume region (P) and the upper laminar transport ribbon (R), see Fig.~\ref{fig:pearson_spatial}.

\subsection{Flow network construction}

Different from the previous example, we follow the ``classical'' approach of empirically estimating pairwise correlation coefficients between the simulated nutrient time series at each grid point as the basis for flow network construction. In order to eliminate spurious correlations due to spatial auto-correlations and to mimic the effect of observational noise in real-world geophysical data sets, standard Gaussian white noise $\epsilon_m(t) \ll \hat{x}_m(t)$ with zero mean and unit variance is first added to the original time series $\hat x_m(t)$:
\begin{equation}
	x_m(t) =  \hat x_m(t) + \epsilon_m(t). 
\end{equation}
\noindent
Notably, this stochastic perturbation is about 1-2 orders of magnitude smaller than the typical range of nutrient concentrations (cf.~Fig.~\ref{fig:plankton_sample}). The resulting noisy time series $x_m(t)$ are normalized to obtain records with zero mean and unit variance at each grid point. 

Subsequently, we use the thus obtained data set for constructing a flow network with $N=6,161$ vertices. Here, each vertex again represents a grid point and its associated time series of normalized nutrient concentrations. Statistical similarity between the time-evolution at two grid points $m$ and $n$ is measured by means of the lagged Pearson cross-correlation
\begin{equation}
	C_{mn} = \max_{\tau\in[-T_C,0[}\; \langle x_m(t) x_n(t-\tau)\rangle.
	\label{eq:ccflag}
\end{equation}
\noindent
Although it does not account for nonlinear interdependencies between time series, Pearson correlation was chosen here because of its lower computational costs (and higher robustness of estimates from short time series) in comparison with potential alternative measures. Moreover, Pearson correlations are most commonly utilized in current studies on climate networks. However, in the example studied here, the obtained results do not change markedly when using other similarity measures like Spearman's Rho or Kendall's Tau (not shown). The choice of a small negative time lag $\tau$ makes the resulting network directed and ensures the correct edge direction. 

The above description of our example setting highlights three important differences in comparison with the first model system studied in Section~\ref{sec:results}: (i) We have to cope with a non-stationary (more specifically, periodic) velocity field rather than a stationary one. (ii) The overall complexity of the system is larger. (iii) We study correlations reflecting the non-trivial deterministic dynamics of a scalar variable instead of such between exclusively stochastic fluctuations, which are relatively weak in the second example due to the low magnitude of the imposed noise. As a fourth distinctive difference, in the following, we consider an \emph{unweighted} network representation. For this purpose, we select an $\alpha$-percentile of the empirical distribution of all non-diagonal elements $C_{mn}$ ($m\neq n$) of $\mathbf{C}$, denoted as $C^*$, and set
\begin{equation}
a_{mn}=\Theta(C_{mn}-C^*)-\delta_{mn},
\label{def:network_binary}
\end{equation}
\noindent
where $\Theta(\cdot)$ denotes the Heaviside function and $\delta_{mn}$ is again Kronecker's delta. $\mathbf{A}=(a_{mn})$ represents the adjacency matrix of the resulting \emph{flow network}, capturing the ``statistical backbone'' of the underlying velocity field based on the scalar observables $x_m$. A corresponding thresholded correlation approach is widely used for constructing \emph{functional networks} from spatio-temporal data sets in a variety of disciplines ranging from climatology \cite{Tsonis:2004} over neurosciences \cite{Park2013} to economics \cite{Boginski2005}.

\begin{figure}
	\includegraphics[width = 1.0\columnwidth]{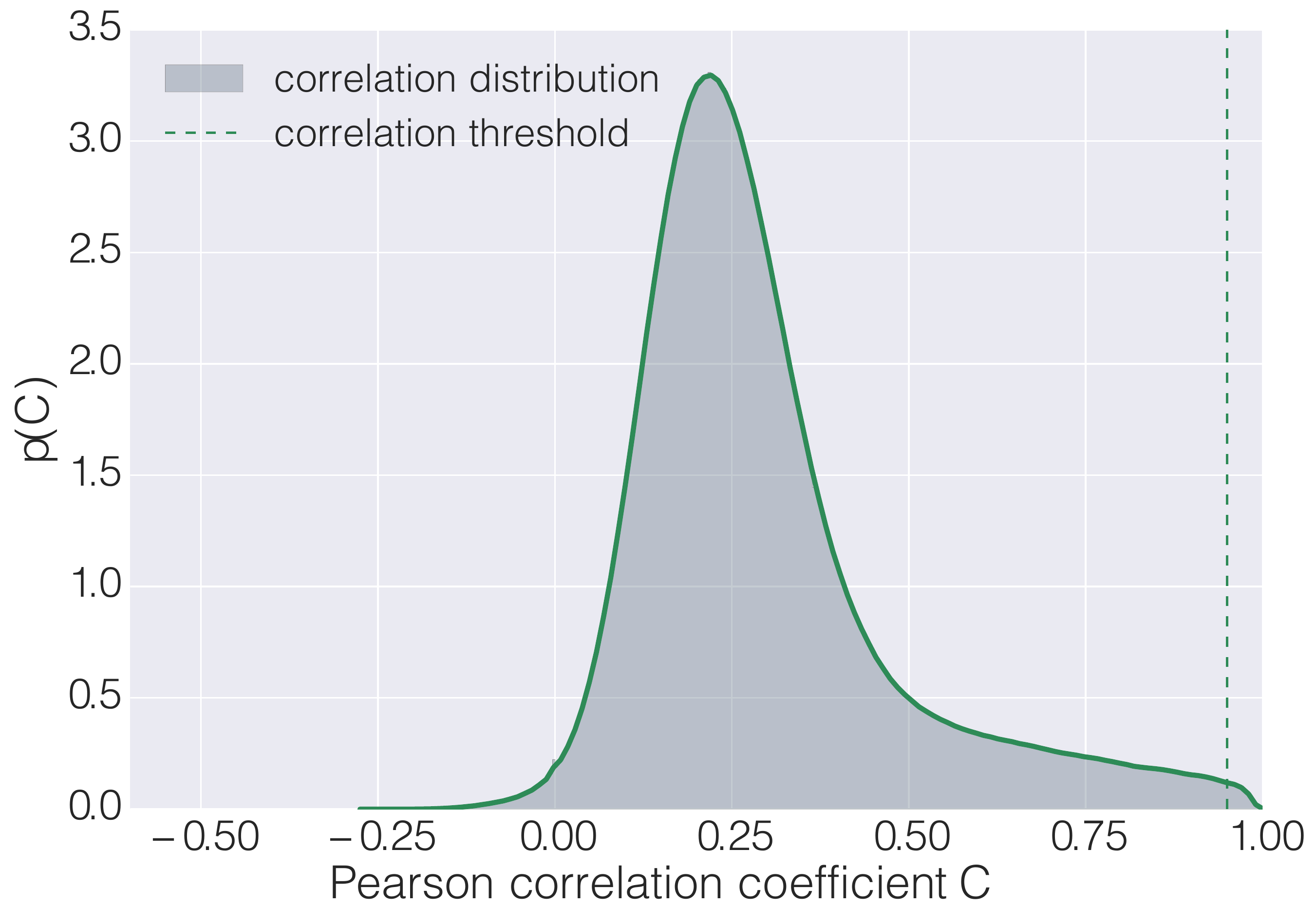}
	\caption{Empirical distribution $p(C)$ of correlation values for the complex flow example. The solid line shows a kernel density estimate; the dashed line represents the minimum correlation threshold of $C^* = 0.95$ used for the flow network construction.}
	\label{fig:pearson}
\end{figure}

Figure~\ref{fig:pearson} shows the obtained distribution of lagged maximum correlation values. It can be recognized that this distribution has marked positive skewness and a unimodal shape with a mode at $C\approx 0.2$ that originates from the majority of noisy and pairwise at most weakly correlated -- or in several cases even practically uncorrelated -- time series. The latter are neglected in the network construction by employing a correlation threshold of $C^*=0.95$. Thereby, we derive a directed network (since commonly $C_{mn}\neq C_{nm}$) with an edge density of $\rho=K/N(N-1)=0.0036$ (implying that $C^*$ corresponds to the 99.64\% percentile of the empirical distribution of all pairwise correlation values). Clearly, the absolute number of edges depends on the distribution of correlation values $C_{mn}$ and the specific threshold $C^*$. For the observed right-tailed distribution of correlation values (Fig.~\ref{fig:pearson}), sufficiently high $C^*\gtrsim 0.8$ guarantee qualitatively robust flow network patterns. 

\subsection{Flow network characteristics}

Even without the construction of a flow network, we can already visually identify the previously mentioned three distinct regions of interest (ROIs) from the spatial patterns of mean correlations (Fig.~\ref{fig:pearson_spatial}). In the following, we will further characterize the flow regimes within these three regions by means of our network measures, thereby interpreting the correlation structure underlying the flow network more thoroughly. In addition to the resulting spatial patterns of degree and local anisotropy, we also employ three complementary topological flow network characteristics widely used in complex network studies across disciplines \cite{Boccaletti:2006a}:
\begin{itemize}
\item betweenness $B_m$, which measures the fraction of shortest paths in the network that traverse a vertex $m$,
\item local clustering coefficient $c_m$, giving the probability of vertices connected with $m$ to be mutually connected among themselves, and
\item mean edge length $\bar{l}_m$, quantifying the average spatial distance covered by the edges starting at $m$.
\end{itemize}
\noindent
In what follows, for the topological network characteristics we will restrict ourselves to the discussion of undirected network measures -- despite the fact that the constructed flow network is directed. Specifically, we consider each edge in the network to be bidirectional if it is present in at least one direction. There are two reasons for making the corresponding simplification: On the one hand, there is no unique local clustering coefficient for directed networks; instead, one has to distinguish between different motifs composed of three vertices \cite{Fagiolo2007}. On the other hand, at least for the degrees we find a rather large correlation between in-degree and out-degree fields ($r\approx 0.82$) indicating that both properties reveal at least qualitatively similar information. Note that this is distinctively different for the anisotropy, where the values of $R_m^{out}$ for the directed and $R_m$ for the undirected flow network hardly show any statistically relevant correlation. Therefore, we will specifically consider the out-anisotropy as the geometric measure of choice. A more detailed comparison between directed and undirected geometric characteristics will be a subject of future work.

For a concise overview, the main results of our analysis are briefly summarized in Tab.~\ref{tab:PN_results}. This synthesis shows that the three dynamical regions P, U and R exhibit unique features when considering combinations of the selected network measures. None of the measures alone is sufficient for obtaining a classification of the corresponding regions and associated transport regimes. In turn, the latter task can only be achieved when the different measures are combined. 

\begin{table}[thb]
	\centering
	\begin{tabular}{lccc}
		\hline
		ROI								& plumes		& upwelling				& 	ribbon		\\
		\hline\hline            		
		range in $x^{(1)}$						& $2..6$		& $1..3$				& 	$1..5$			\\
		range in $x^{(2)}$						& $-1..1$		& $1..2$				& 	$\approx 2$		\\
		mean edge length $\bar{l}_m$ 	& low			& very low 				& 	high			\\
		degree $k_m$ 					& high 			& moderate				& 	high			\\
		betweenness $B_m$ 				& various		& very high				& 	very high		\\
		clustering coefficient $c_m$ 				& high 			& moderate				& 	moderate		\\
		out-anisotropy $R_m^{out}$			& various		& moderate				& 	high			\\
		\hline
	\end{tabular}
		\caption{Synthesis of typical ranges of local network measures inside the three distinct regions plumes (P), upwelling (U) and ribbon (R) of the flow network constructed from nutrient concentrations. In order to support the visual analysis of spatial patterns of network characteristics in the different ROIs (Fig.~\ref{fig:plankton_results}), a classification is provided in qualitative terms (\emph{very low}, \emph{low}, \emph{moderate}, \emph{high}, \emph{very high}, as well as \emph{various} in case of more ambivalent values).}
	\label{tab:PN_results}
\end{table}

\begin{figure*}[htbp]
	\centering
		\includegraphics[width=0.48\textwidth,height=5.5cm]{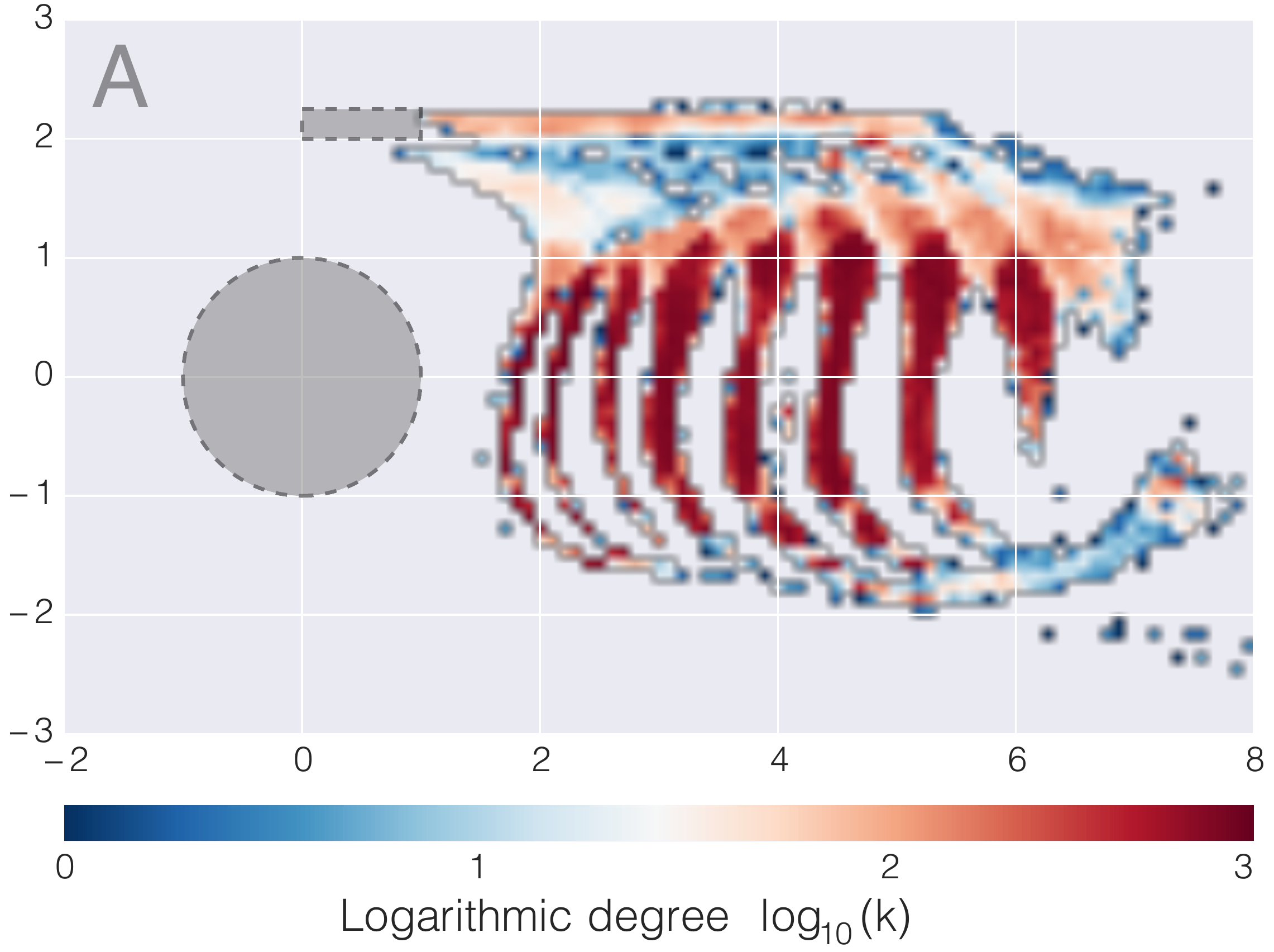} \hfill
		\includegraphics[width=0.48\textwidth,height=5.5cm]{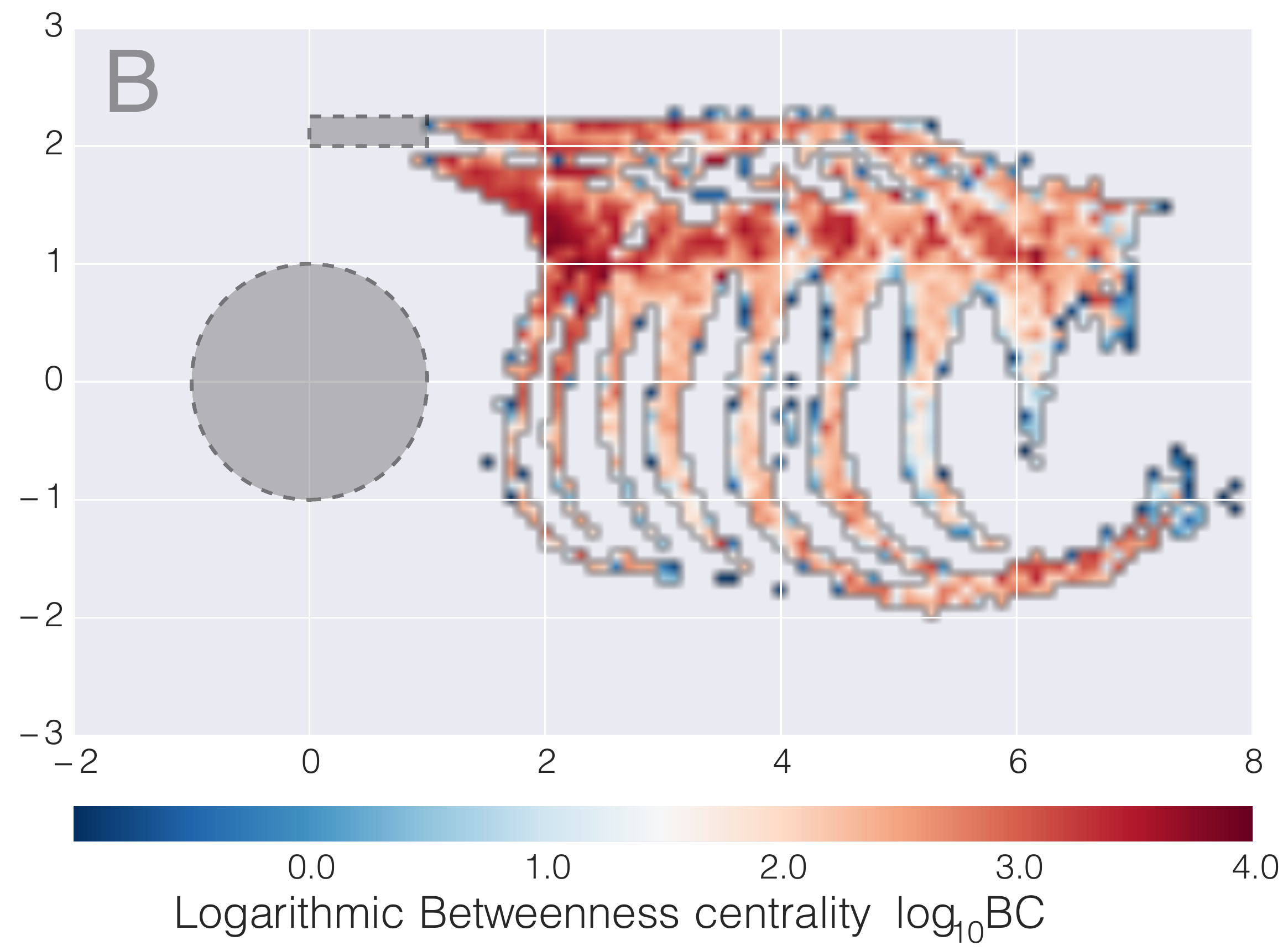} \\
		\includegraphics[width=0.48\textwidth,height=5.5cm]{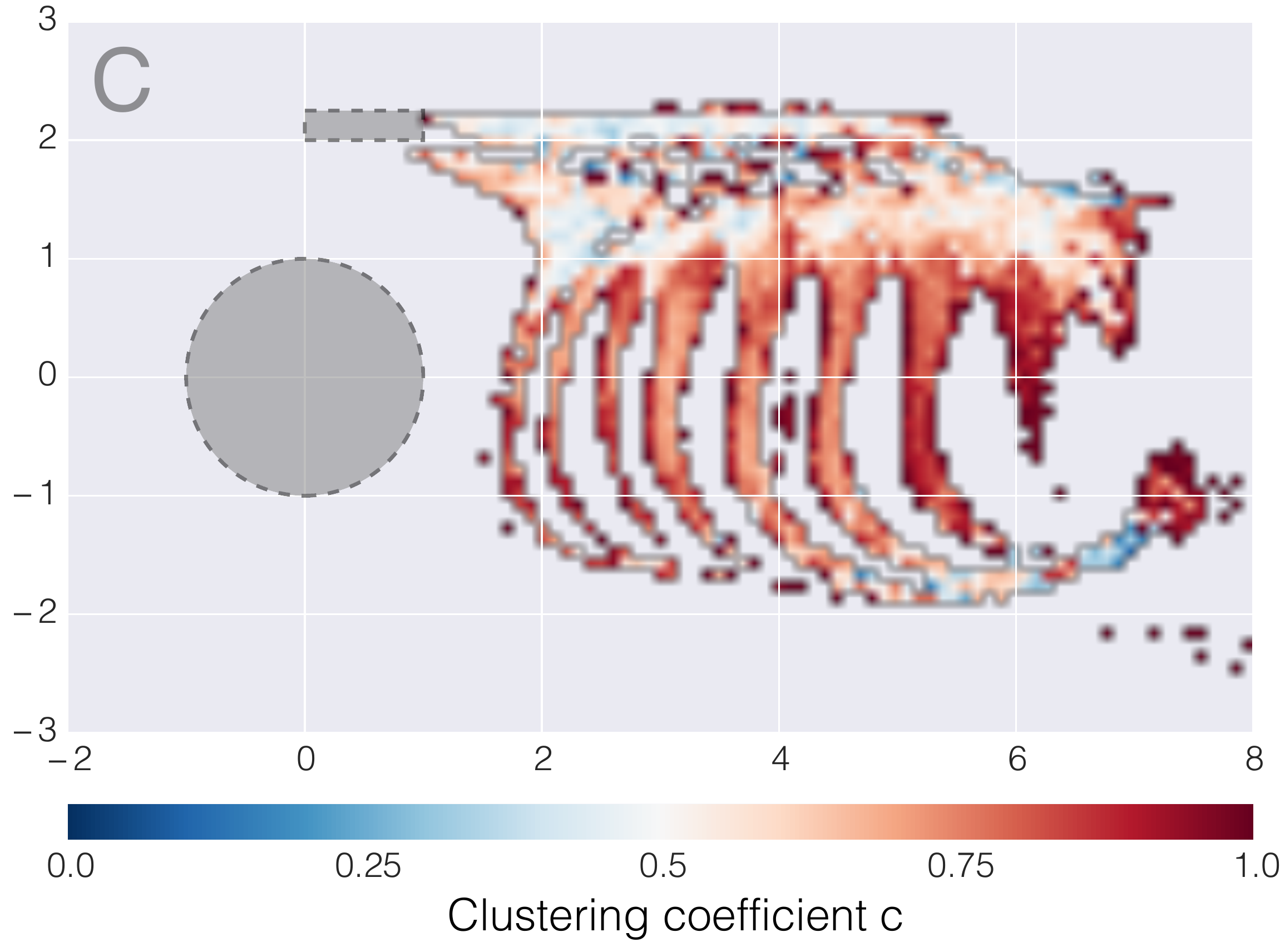} \hfill
		\includegraphics[width=0.48\textwidth,height=5.5cm]{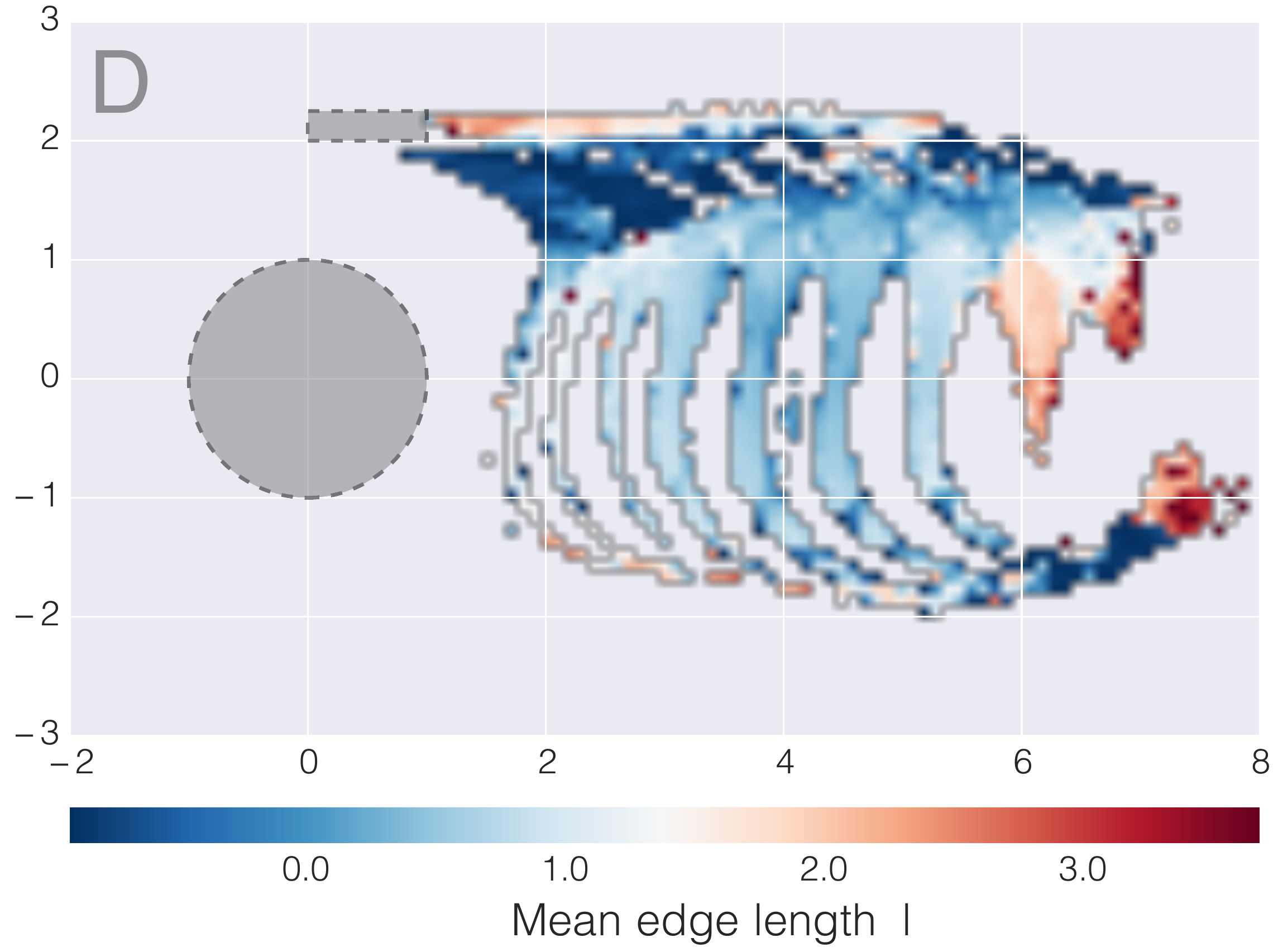} \\
		\includegraphics[width=0.48\textwidth,height=5.5cm]{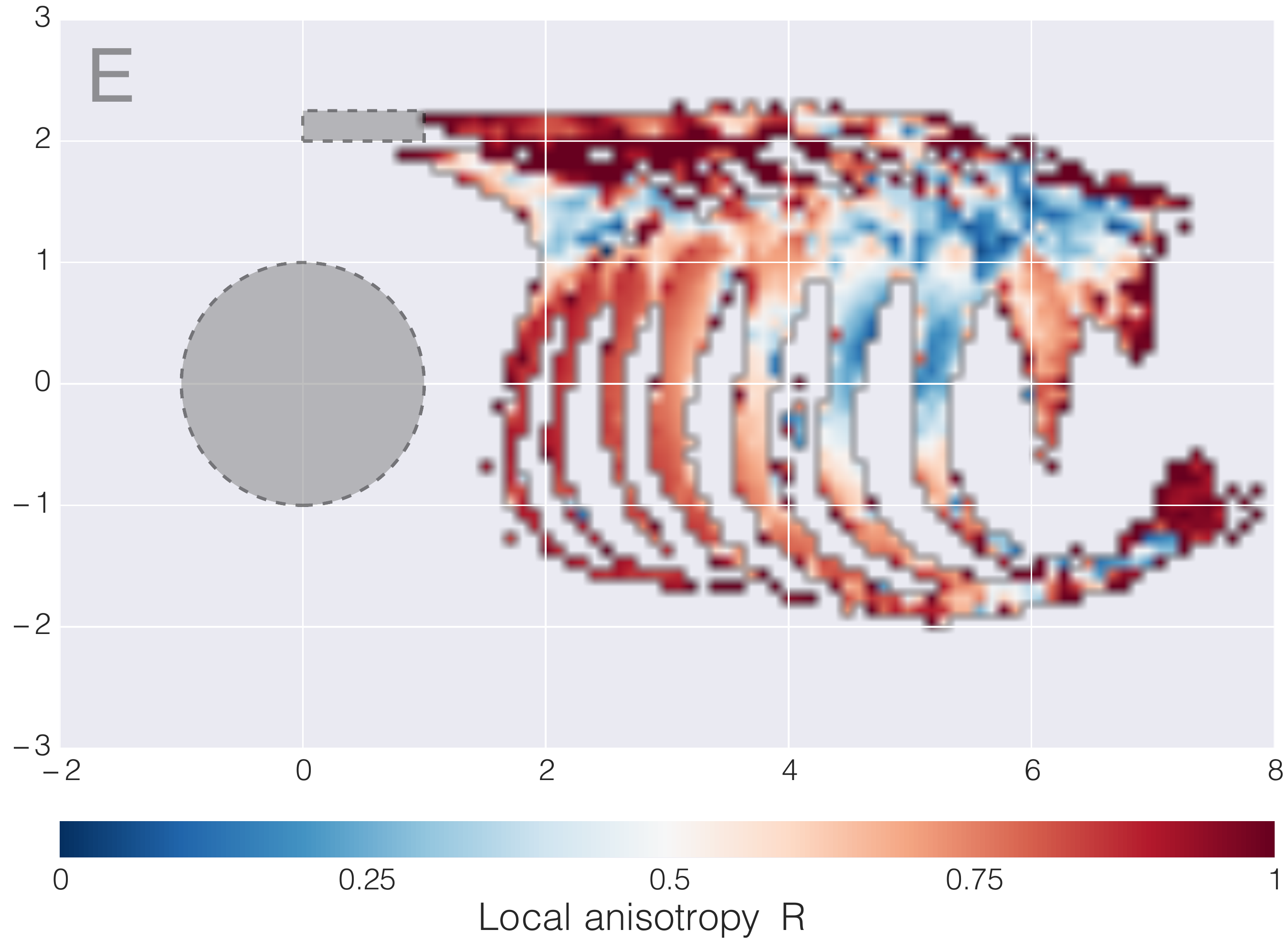}
	\caption{(Color online) Local properties of the complex flow network: (A) (logarithm of) degree, (B) (logarithm of) betweenness, (C) local clustering coefficient, (D) mean edge length and (E) local out-anisotropy. For the interpretation of the measures as well as their meaning in combination, see the main text. A concise overview is given in Tab.~\ref{tab:PN_results}.}
	\label{fig:plankton_results}
\end{figure*}

Our combined analysis of the spatial patterns displayed by the different topological and geometric network properties (see Fig.~\ref{fig:plankton_results}) reveals a detailed picture of the inherent correlation structure and its resulting signatures in the flow network. In this context, note that for the applied very high correlation threshold $C^*$, the flow network contains a large number of isolated vertices outside the regions U, P and R. In the following, we will only consider those parts of the network that are actually connected under the given setting.

The surrounding of the upwelling region (U) exhibits intermediate values of degree, local clustering coefficient as well as local anisotropy. In combination with very low values of the mean edge length, the corresponding part of the flow network is characterized by a rather localized connectivity with transport patterns of not strongly focused directionality. Accompanied by very high values of betweenness centrality in this region, a local and short-range transport regime is indicated. This interpretation agrees with the high nutrient concentration being slowly released from the upwelling region (indicated by the rectangular shape in Fig.~\ref{fig:pearson_spatial}) and fed into either the vortex region in the wake of the island or the ribbon region before being advected out of the study area.

The central plumes region (P) shows low values of mean edge lengths but very high degrees and high local clustering coefficients, pointing to a dense and spatially localized connectivity~\cite{Radebach:2013}. In combination with a broad range of betweenness and anisotropy values, this indicates a relatively slow propagation of the nutrient concentration patterns. Areas of higher anisotropy along the upstream border of the plumes (Fig.~\ref{fig:plankton_results}E) probably reflect that there are hardly any strong correlations with vertices that are even closer to the circular obstacle. In turn, given the consistently low edge length (Fig.~\ref{fig:plankton_results}D), a strong correlation with the other parts of the plume in the successive time steps is plausible. A sole effect of spatio-temporal autocorrelations can be practically ruled out due to the consideration of possible time lags. Moreover, we observe low local anisotropy values in the region between transport ribbon and plumes, which are exactly those regions where the flow changes its direction sharply. This observation matches well the corresponding results for the more basic meandering flow pattern studied in Section~\ref{sec:results}. In a similar way, the lower anisotropy values at the downstream end of the plume region probably reflect the fact that due to the periodicity of the flow, in this region there exist strong correlations with vertices in various different directions, but with possibly different lags (note again that we did not consider lag-zero correlations here, but explicitly allow for certain time delays).

Finally, the upper transport ribbon (R) is distinguished by a strong and long-ranging transport regime of relatively laminar character. This is indicated by high values of mean edge length and degree, as well as a strong directionality, suggested by high anisotropy values and very high values of betweenness. The local clustering coefficient in this region is bound to moderate values, which supports the interpretation of a relatively small number of interconnections with other parts of the flow and, thus, a comparably straight connectivity pattern.

\begin{figure*}
	\centering
	\includegraphics[width = 0.85\textwidth]{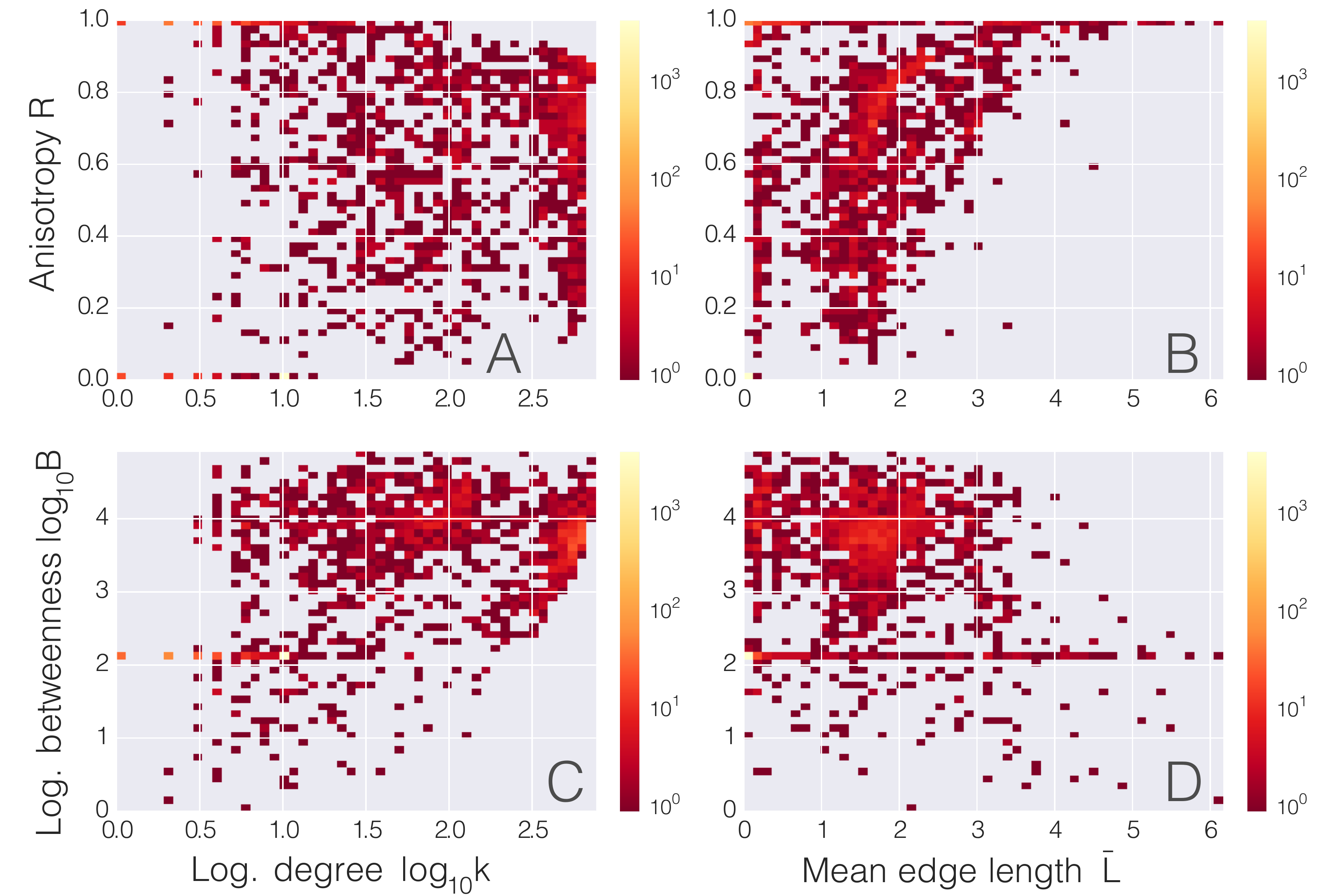}
	\caption{(Color online) Pairwise joint probability distributions of selected network measures in the complex flow network. Colors again indicate the frequency of combinations of the values of the different measures.}
	\label{fig:plankton_joint_prob_plot}
\end{figure*}

The classification of different transport regimes coinciding with regions of vertices with distinguished properties is further underpinned by the joint probability densities of pairs of network characteristics displayed in Fig.~\ref{fig:plankton_joint_prob_plot}A-D. For example, the plumes region (P) is represented by a group of many vertices with high degree and a broad range of betweenness values (Fig.~\ref{fig:plankton_joint_prob_plot}C). At the same time, this group exhibits a broad range of edge anisotropy values (Fig.~\ref{fig:plankton_joint_prob_plot}A) and relatively low values of mean edge length (Fig.~\ref{fig:plankton_joint_prob_plot}D). In a similar way, the two other ROIs are associated with different groups of vertices visible in Fig.~\ref{fig:plankton_joint_prob_plot}.

In summary, by combining the different structural and geometric aspects, the dynamical features of the system encoded in the correlation structure can be recovered. During the analysis, edge anisotropy contributed a so far missing aspect to the identification and differentiation of the distinct regions and helped to point out different dynamical regimes of transport in these regions.

\section{Conclusions and Outlook}
\label{sec:conclusion_outlook}

Complex networks embedded in some physical space are often only partly described by their topological characteristics. Specifically, it is known that many classical network properties (focusing exclusively on the mutual linkage between vertices) are strongly predetermined by the spatial positions of vertices and edges~\cite{Bialonski:2010,Wiedermann2015}. Examples for this phenomenon include climate networks \cite{Rheinwalt:2012,Radebach:2013} and brain networks \cite{Bialonski:2010}. Taking this additional information into account, a more holistic picture of the system's structural organization can be drawn. 

In the context of geometric network properties, previous attention has mostly focused on the edge length distributions and ``trend'' (edge orientation) entropies, the latter being exclusively used in the field of road network analysis so far \cite{Gudmundsson:2013,Mohajeri:2013,Mohajeri:2014}. Beyond the latter concepts, the present work has introduced a new measure for quantifying the anisotropy of edges adjacent to a given vertex in a spatial network. The proposed approach can be generally applied to characterizing the spatial structure of networks in a variety of fields. One prominent potential application are road networks \cite{Lammer:2006,Chan:2011} or, more generally, infrastructures, where cost-optimization typically calls for explicit consideration of spatial constraints when building the network \cite{Gastner:2006,Kreher:2012}. 

In the present work, we have restricted our attention to flow networks embedded in a two-dimensional Euclidean space with Cartesian coordinates. Notably, there exist also numerous examples of spatial networks naturally having a three-dimensional structure, including network representations of the Earth's atmosphere~\cite{Donges:2011a}, ocean currents~\cite{Feng2014}, the human brain~\cite{Bialonski:2010}, intracellular transport or technical flow systems \cite{Helbing2009}. The framework proposed here is general enough to be directly applicable to such networks as well. Moreover, in the case of non-Euclidean geometries \cite{Krioukov:2012}, one can extend the local and global anisotropy characteristics defined in this work by utilizing concepts from differential geometry, i.e., incorporating the curvature tensor of the respective metric space.

By applying the novel concept of local anisotropy to two correlation-based flow networks constructed in different ways, this paper has contributed to the ongoing development of a new correspondence principle for the investigation of spatially embedded networks representing dynamical systems. In combination with classical topological network characteristics like vertex degree, the local anisotropy facilitates the identification of macroscopic regions that exhibit directed flow and, hence, transport. We have demonstrated the potential usefulness of our approach for two physically well-understood prototypical model flows of different size and complexity. 

One main potential field of application of correlation-based (functional) flow networks as studied here is qualitatively deducing information about the hard to observe velocity field of geophysical flows based on similarity patterns of more directly observable scalar variables like temperature, sea-surface salinity or nutrient or phytoplankton concentrations, where large-scale data sets have recently become available from extensive remote sensing campaigns. Although this approach does not allow for a detailed quantitative reconstruction of the flow itself, by jointly assessing conceptually different topological and geometric properties of flow networks, relevant flow structures can be identified and possibly attributed. Specifically, different domains of the flow (e.g., advection versus diffusion-dominated areas) are characterized by different combinations of values of these network measures. This aspect becomes especially important when investigating less well-understood systems and phenomena, where the detailed physical description and understanding is still subject of ongoing research. In this spirit, our results may provide a basis for gaining a better understanding of the spatio-temporal organization of a broad variety of complex systems, including possible applications to climate, human neuro-physiology or transportation systems.

We emphasize that the consideration of complementary aspects for exploring unknown phenomena in some data-driven way commonly provides a more detailed picture than focusing on individual measures. 
Notably, this statement is supported by other recent studies on flow networks \cite{Molkenthin:2013a, Tupikina2015}, where no unique correspondence between the values of commonly studied topological network measures like degree or betweenness and the underlying flow structures could be found. 

One aspect neglected by the correlation-based approach followed in this as well as many other recent contributions to this field is that the consideration of dynamical similarities disregards available more detailed information on the temporary dynamics, 
which would be particularly relevant in case of non-stationary flows commonly observed in geophysical systems like the atmosphere and oceans \cite{Watts1982} or in neuroscience. More detailed studies of the latter cases call for alternative approaches such as a time-resolved analysis \cite{Radebach:2013} or coarse-graining the dynamics and studying it within some Markovian framework, where networks are constructed based on transition probabilities between discrete ``states'' \cite{Mieruch:2010}. A detailed exploration of the latter type of approach will be subject of future research.

\subsection*{Acknowledgements}
This work was financially supported by the German Research Foundation (DFG) via the DFG Graduate School 1536 (``Visibility and Visualization''), the European Commission via the Marie-Curie ITN LINC (P7-PEOPLE-2011-ITN, grant no.~289447), the German Federal Ministry for Education and Research (BMBF) via the BMBF Young Investigator's Group CoSy-CC$^2$ (``Complex Systems Approaches to Understanding Causes and Consequences of Past, Present and Future Climate Change, grant no.~01LN1306A'') and the project GLUES, the Stordalen Foundation (via the Planetary Boundary Research Network PB.net), the Earth League's EarthDoc program, and the Volkswagen Foundation via the project ``Recurrent extreme events in spatially extended excitable systems: Mechanism of their generation and termination'' (grant no.~85391). The presented research has greatly benefited from discussions with Emilio Hern\'andez-Garcia and Crist\'obal L\'opez. Parts of the network calculations have been performed using the Python package \texttt{pyunicorn} \cite{Donges2015} (see \url{http://tocsy.pik-potsdam.de/pyunicorn.php}). \texttt{pyunicorn} is freely available for download at \url{https://github.com/pik-copan/pyunicorn}.

\bibliographystyle{unsrt}
\bibliography{bibliography,hannes_nora_liuba_additional}

\end{document}